\documentclass[10pt,aps,prd,nofootinbib,superscriptaddress,twocolumn]{revtex4}
\usepackage[utf8]{inputenc}
\DeclareUnicodeCharacter{200B}{{\hskip 0pt}}
\usepackage{color}
\usepackage{graphicx}
\usepackage{amsmath}
\usepackage{amssymb}
\usepackage{bm}
\usepackage{acronym}
\usepackage{ifthen}
\usepackage{blindtext}
\usepackage[normalem]{ulem}
\usepackage{hyperref}
\usepackage{etoolbox}
\usepackage{fancyhdr}
\usepackage{xspace}
\usepackage{textcomp}
\usepackage{multirow}
\usepackage[caption=false]{subfig}
\usepackage{lineno}
\usepackage{tabularx}
\hypersetup{
    colorlinks=true,
    linkcolor=blue,
    filecolor=magenta,      
    urlcolor=black,
		citecolor=red,
		}
\begin{document}

\title{Analytical and spectral characterization of floral patterns in
higher-order polynomial pp-waves}

\author{F. L. Carneiro}\email{fernandolessa45@gmail.com}
\affiliation{Universidade Federal do Norte do Tocantins, 77824-838, Aragua\'ina, TO, Brazil}

\author{H. P. de Carvalho}
\email{hyagopeixoto266@gmail.com}
\affiliation{Universidade Federal do Norte do Tocantins, 77824-838, Aragua\'ina, TO, Brazil}

\author{L. A. Cabral}\email{luis.cabral@ufnt.edu.br}
\affiliation{Universidade Federal do Norte do Tocantins, 77824-838, Aragua\'ina, TO, Brazil}

\author{M. P. Lobo}
\email{matheus.lobo@ufnt.edu.br
}
\affiliation{Universidade Federal do Norte do Tocantins, 77824-838, Aragua\'ina, TO, Brazil}

\begin{abstract}
We investigate analytically and numerically the floral deformation of
particle rings induced by higher-order polynomial modes of pp-waves.  The
transverse tidal eigendirections determine $m$ interlaced sectors of radial
stretching and compression, while a weak-pulse solution provides the complete
first-order response of the ring.  Direct geodesic integration for $m=2,3,4$
confirms the predicted orientation and discrete symmetry.  A Fourier analysis
shows that the response remains overwhelmingly dominated by the angular mode
$n=m$; in particular, the loops found for $m=4$ arise from geometrical folding
of the fundamental mode rather than strong harmonic mixing.  The temporal
moments of the pulse further distinguish displacement from velocity memory
without changing this angular signature.

\end{abstract}


\maketitle

\date{\today}

\section{Introduction}

Gravitational memory arises in radiative solutions of Einstein equations and describes a persistent change in the relative configuration or motion of freely falling particles after the passage of gravitational radiation \cite{christodoulou1991nonlinear,thorne1992gravitational}. Some aspects of this effect are particularly transparent for massive particles interacting with sandwich plane gravitational waves (GWs), for which spacetime becomes flat in the asymptotic regions, as well as in the impulsive limit \cite{podolsky2014gyratonic,zhang2018memory}.

In recent literature, Gaussian-like pulses are often considered and the geodesic equations are solved numerically for massive particles \cite{zhang2017memory,maluf2018plane}, while memory may also persist beyond the sandwich approximation \cite{zhao2026plane,datta2024memory}. The initial separation between particles can then be permanently altered, with or without a change in their asymptotic relative velocity depending on the pulse parameters \cite{zhao2026approximate}. The latter case is usually referred to as the \textit{displacement memory effect} (DM), whereas a nonvanishing change in the asymptotic relative velocity characterizes the \textit{velocity memory effect} (VM) \cite{zhang2024displacement}. Ben Achour and Uzan~\cite{achour2024displacement} classified the conditions for velocity and displacement memory in vacuum plane waves in terms of the wave profile and the initial relative motion, explicitly demonstrating the possibility of displacement memory.


The parameters of the pulse have a significant impact on the occurrence of DM. For several profiles, DM appears only for specific values of the wave parameters \cite{zhang2025displacement,elbistan2026globally,zhang2025flyby,achour2024displacement}. This observation is interesting for pp-waves written in Brinkmann coordinates as
\begin{equation}\label{line_element}
ds^{2}=H(u,x,y)\,du^{2}+dx^{2}+dy^{2}-2\,du\,dv,
\end{equation}
with
\begin{equation}\label{einstein_equation}
(\partial_{x}^{2}+\partial_{y}^{2})H(u,x,y)=0,
\end{equation}
propagating along the null direction $v$, transverse to the flat space spanned by $x$ and $y$. Consider, for instance,
\begin{equation}
H(u,x,y)=A\,f(u)\,\Phi(x,y).
\end{equation}
Under the null-coordinate rescaling $\bar u=\lambda u$ and
$\bar v=v/\lambda$, the Brinkmann form is preserved while $A f(u)$ is mapped
to $A\lambda^{-2}f(\bar u/\lambda)$.  The amplitude and longitudinal width are
therefore not independent, and any special memory-producing value of $A$ must
be specified together with the pulse scale~\cite{zhang2026memory}.


The asymptotic behavior of the pulse is also essential. Zhao and Cao recently showed that the asymptotic class of particle motion is controlled by the decay of the profile tail, deriving weighted decay criteria for asymptotic freedom and expressing them in terms of the accumulated tidal matrix \cite{zhao2026decay}. More generally, memory is a cumulative effect of the interaction history, since the asymptotic relative velocity depends on the tidal field integrated along the null evolution \cite{carneiro2026memory}.

Most studies of plane-wave memory have focused on the quadrupolar solutions of Eq.~\eqref{einstein_equation}, corresponding to the usual $+$ and $\times$ polarizations. More generally, the two-dimensional Laplace equation admits the polynomial modes
\begin{equation}
\Phi_{mR}(x,y)=\Re[(x+iy)^{m}],
\,
\Phi_{mI}(x,y)=\Im[(x+iy)^{m}],
\end{equation}
with the $+$ and $\times$ modes corresponding to the real and imaginary parts for $m=2$. Although higher-order polynomial modes have recently attracted attention in the study of geodesic dynamics \cite{rossetto2026wada}, their memory properties remain comparatively unexplored. In Ref.~\cite{carneiro2026memory}, we investigated VM and the associated exchange of energy between particles and the wave for these higher modes. An initially circular ring of particles was found to develop an $m$-dependent floral pattern consisting of a characteristic number of loops.

Although this behavior was obtained numerically in Ref.~\cite{carneiro2026memory}, its angular structure admits a simple analytical explanation. The transverse curvature defines a tidal matrix whose eigenvectors determine the principal directions of stretching and compression, while its eigenvalues determine their intensities. In this Note, we use this structure to explain analytically the floral patterns generated by higher-order polynomial modes and their dependence on the mode number $m$.

\section{Tidal structure of higher-order polynomial modes}\label{sec2}

We now derive the transverse tidal structure associated with a single
polynomial mode.  Introducing polar coordinates
$x=r\cos\varphi$ and $y=r\sin\varphi$, a real mode with an arbitrary angular
orientation can be written as
\begin{equation}\label{higher_mode_profile}
 H_m(u,r,\varphi)=A_m f(u)r^m
 \cos\!\left[m(\varphi-\varphi_0)\right],
 \qquad m\geq2.
\end{equation}
We take $f(u)$ to be dimensionless, so that $A_m$ has dimensions of
${\rm length}^{-m}$ and $H_m$ is dimensionless.  The choices $\varphi_0=0$ and
$\varphi_0=\pi/(2m)$ give, respectively, the real and imaginary parts of
$(x+iy)^m$.

For the metric convention in Eq.~\eqref{line_element}, the transverse
geodesic equations, with $u$ used as an affine parameter, are
\begin{equation}\label{transverse_geodesics}
 \frac{d^2x^i}{du^2}=\frac{1}{2}\partial_iH_m,
 \qquad x^i=(x,y).
\end{equation}
Accordingly, the relative acceleration of neighboring geodesics is governed
by the symmetric matrix
\begin{equation}\label{tidal_matrix_definition}
 {\cal T}_{ij}=\frac{1}{2}\partial_i\partial_jH_m,
 \qquad
 \frac{d^2\xi^i}{du^2}={\cal T}^{i}{}_{j}\xi^j.
\end{equation}
This equation fixes our sign convention for the tidal matrix.  Reversing the
curvature convention changes the overall sign of ${\cal T}$ and interchanges
the labels of its two principal directions, without changing the geometry
derived below.

It is useful first to evaluate Eq.~\eqref{tidal_matrix_definition} in
Cartesian coordinates.  Defining
\begin{equation}
 \alpha_m=(m-2)\varphi-m\varphi_0
\end{equation}
and
\begin{equation}\label{sigma_definition}
 \sigma(u,r)=\frac{A_m f(u)}{2}
 m(m-1)r^{m-2},
\end{equation}
direct differentiation gives
\begin{equation}\label{tidal_cartesian}
 {\cal T}_{(x,y)}=\sigma
 \begin{pmatrix}
  \cos\alpha_m & -\sin\alpha_m\\
  -\sin\alpha_m & -\cos\alpha_m
 \end{pmatrix}.
\end{equation}
The vanishing trace is the local manifestation, in the transverse plane, of
the vacuum equation \eqref{einstein_equation}.

The geometrical interpretation is clearest in the orthonormal polar basis
$(\widehat{\bm e}_r,\widehat{\bm e}_\varphi)$.  This distinction is important:
the coordinate vector $\partial_\varphi$ has norm $r$ and therefore cannot be
used directly as a unit direction.  If
\begin{equation}
 Q(\varphi)=
 \begin{pmatrix}
  \cos\varphi & -\sin\varphi\\
  \sin\varphi & \cos\varphi
 \end{pmatrix}
\end{equation}
maps orthonormal polar components to Cartesian components, then
${\cal T}_{(r,\varphi)}=Q^{T}{\cal T}_{(x,y)}Q$.  Equation
\eqref{tidal_cartesian} consequently becomes
\begin{equation}\label{tidal_polar}
 {\cal T}_{(r,\varphi)}=\sigma
 \begin{pmatrix}
  \cos\chi & -\sin\chi\\
  -\sin\chi & -\cos\chi
 \end{pmatrix},
 \qquad
 \chi=m(\varphi-\varphi_0).
\end{equation}
Thus the magnitude of the tidal field and the orientation of its principal
axes separate naturally: all radial and temporal dependence is contained in
$\sigma$, whereas all angular dependence is contained in $\chi$.

The matrix multiplying $\sigma$ in Eq.~\eqref{tidal_polar} squares to the
identity. Therefore, the eigenvalues of the resulting matrix ${\cal T}$ are
\begin{equation}\label{tidal_eigenvalues}
 \lambda_+=\sigma,
 \qquad
 \lambda_-=-\sigma,
\end{equation}
with corresponding unit eigenvectors
\begin{align}
 \widehat{\bm e}_+
 &=\cos\!\left(\frac{\chi}{2}\right)\widehat{\bm e}_r
   -\sin\!\left(\frac{\chi}{2}\right)\widehat{\bm e}_\varphi,
 \label{positive_eigenvector}\\
 \widehat{\bm e}_-
 &=\sin\!\left(\frac{\chi}{2}\right)\widehat{\bm e}_r
   +\cos\!\left(\frac{\chi}{2}\right)\widehat{\bm e}_\varphi.
 \label{negative_eigenvector}
\end{align}
Since a principal axis is unchanged when its eigenvector is multiplied by
$-1$, its angle is defined modulo $\pi$.  In particular, the angle made by
$\widehat{\bm e}_+$ with the local radial direction is
\begin{align}
 \beta_+&=-\frac{m}{2}(\varphi-\varphi_0)\pmod{\pi},
 \label{principal_angle}\\
 \beta_-&=\beta_++\frac{\pi}{2}\pmod{\pi}.
\end{align}
For $\sigma>0$, $\widehat{\bm e}_+$ and
$\widehat{\bm e}_-$ are the instantaneous stretching and compression axes,
respectively; when $\sigma<0$, these roles are interchanged.

Several checks follow immediately from this result:
\begin{equation}\label{tidal_checks}
 \operatorname{tr}{\cal T}=0,
 \qquad
 \det{\cal T}=-\sigma^2,
 \qquad
 \operatorname{tr}({\cal T}^2)=2\sigma^2.
\end{equation}
For $m=2$, the tidal magnitude is independent of $r$, as expected for the
usual quadrupolar plane wave.  The formal $m=1$ mode has a vanishing Hessian
and produces no tidal curvature.  For every $m>2$, the magnitude scales as
$r^{m-2}$ and vanishes on the propagation axis; at that point the tidal
matrix is zero and its principal directions are not defined.  Away from the
axis, Eq.~\eqref{principal_angle} gives the complete angular evolution of the
principal tidal axes.  Its consequences for a circular distribution of
particles are developed in the following section.

\section{Floral response of particle rings}\label{sec3}

\subsection{Angular sectors and weak-pulse response}\label{sec3_angular}

We first identify the points at which one of the principal tidal axes is
radial.  Suppose initially that $\sigma>0$, so that
$\widehat{\bm e}_+$ is the stretching direction.  From
Eq.~\eqref{principal_angle}, this direction is parallel to
$\widehat{\bm e}_r$ when $\beta_+=0\pmod{\pi}$.  The radial stretching
directions are therefore
\begin{equation}\label{radial_stretching_sectors}
 \varphi_k^{(\mathrm{st})}
 =\varphi_0+\frac{2\pi k}{m},
 \qquad k=0,\ldots,m-1.
\end{equation}
Similarly, $\widehat{\bm e}_-$ is radial at
\begin{equation}\label{radial_compression_sectors}
 \varphi_k^{(\mathrm{co})}
 =\varphi_0+\frac{(2k+1)\pi}{m},
 \qquad k=0,\ldots,m-1.
\end{equation}
Each family thus contains $m$ equally spaced directions, with angular
separation $2\pi/m$.  The stretching and compression families are interlaced
and displaced from each other by $\pi/m$.  If $\sigma<0$, their physical
roles are exchanged, but the two sets of radial directions are unchanged.

The local eigendirections already predict the angular skeleton of the
deformation.  To connect them directly with a ring of finite radius, however,
we use the transverse geodesic equations rather than extrapolate the local
deviation equation across the entire ring.  Resolving
Eq.~\eqref{transverse_geodesics} in the orthonormal polar basis gives
\begin{align}
 \ddot r-r\dot\varphi^2
 &=\frac{1}{2}\partial_rH_m
 =\frac{A_m m}{2}f(u)r^{m-1}\cos\chi,
 \label{radial_geodesic_equation}\\
 r\ddot\varphi+2\dot r\dot\varphi
 &=\frac{1}{2r}\partial_\varphi H_m
 =-\frac{A_m m}{2}f(u)r^{m-1}\sin\chi,
 \label{angular_geodesic_equation}
\end{align}
where a dot denotes differentiation with respect to $u$.  Consider particles
labelled by their initial polar angle $\theta$ and satisfying
\begin{equation}\label{initial_ring}
 \begin{aligned}
  r(u_i,\theta)&=R,&
  \varphi(u_i,\theta)&=\theta,\\
  \dot r(u_i,\theta)&=0,&
  \dot\varphi(u_i,\theta)&=0.
 \end{aligned}
\end{equation}
Evaluating the right-hand sides of
Eqs.~\eqref{radial_geodesic_equation} and
\eqref{angular_geodesic_equation} on the unperturbed ring gives the leading
response,
\begin{align}
 a_r^{(0)}(u,\theta)
 &=\frac{A_m m}{2}f(u)R^{m-1}
   \cos\!\left[m(\theta-\varphi_0)\right],
 \label{initial_radial_acceleration}\\
 a_\varphi^{(0)}(u,\theta)
 &=-\frac{A_m m}{2}f(u)R^{m-1}
   \sin\!\left[m(\theta-\varphi_0)\right].
 \label{initial_tangential_acceleration}
\end{align}
Thus the force has Cartesian angular order $m-1$, as follows from
differentiating a degree-$m$ polynomial, but its radial and tangential
components relative to the initial circle have angular order $m$.

For a weak pulse, Eqs.~\eqref{initial_radial_acceleration} and
\eqref{initial_tangential_acceleration} can be integrated along the
unperturbed trajectories.  Defining
\begin{equation}
 {\cal I}(u)=\int_{u_i}^{u}(u-s)f(s)\,ds,
 \qquad
 {\cal B}_m(u)=\frac{A_m m}{2}R^{m-1}{\cal I}(u),
 \label{response_amplitude}
\end{equation}
one obtains, to first order in the wave amplitude,
\begin{align}
 r(u,\theta)
 &=R+{\cal B}_m(u)
   \cos\!\left[m(\theta-\varphi_0)\right],
 \label{perturbed_ring_radius}\\
 R\left[\varphi(u,\theta)-\theta\right]
 &=-{\cal B}_m(u)
   \sin\!\left[m(\theta-\varphi_0)\right].
 \label{perturbed_ring_angle}
\end{align}
All dependence on the pulse shape, duration, amplitude, and observation time
is contained in $\mathcal{B}_m(u)$; the angular structure is fixed solely by
$m$ and $\varphi_0$.

Equation~\eqref{perturbed_ring_radius} makes the floral pattern explicit.  If
$\mathcal{B}_m>0$, the radial maxima occur at the directions
\eqref{radial_stretching_sectors}, while the minima occur at
\eqref{radial_compression_sectors}.  Consequently, the deformed ring has $m$
outward tips alternating with $m$ inward indentations.  A negative value of
$\mathcal{B}_m$ interchanges maxima and minima and rotates the pattern by
$\pi/m$, without changing their number.  Moreover,
\begin{equation}
 r\left(u,\theta+\frac{2\pi}{m}\right)=r(u,\theta),
 \label{discrete_ring_symmetry}
\end{equation}
so the response has $m$-fold rotational symmetry.  The case $m=2$ is the
familiar elliptic deformation, with two opposite radial maxima.  For $m>2$,
the $m$ maxima become the $m$ petals observed in the numerical
configurations of Ref.~\cite{carneiro2026memory}.  Nonlinear evolution may
change the depth of the indentations or generate additional angular
harmonics, but the discrete symmetry follows from the invariance of both the
profile and the circular initial data under rotations by $2\pi/m$.

This result also clarifies the distinct roles of the tidal eigenvalues and
eigenvectors.  On an initial circle of fixed radius, $|\sigma(u,R)|$ is
independent of $\theta$ and therefore cannot by itself determine the petal
directions.  The orientation comes from the rotating principal axes:
$\widehat{\bm e}_+$ is radial at each outward tip and
$\widehat{\bm e}_-$ is radial at each neighboring indentation, for
$\sigma>0$.  The floral pattern is therefore the global imprint of the
angular rotation of the local tidal eigendirections.

Finally, the imaginary polynomial requires no separate calculation, since
\begin{equation}
 r^m\sin(m\varphi)
 =r^m\cos\!\left[m\left(\varphi-\frac{\pi}{2m}\right)\right].
\end{equation}
It has the same tidal spectrum and the same number of petals as the real
mode, with the entire pattern rigidly rotated by $\pi/(2m)$.

\subsection{Numerical validation and post-pulse memory}\label{sec3_numerical}

We now compare the first-order construction with the direct numerical
integration of Eqs.~\eqref{radial_geodesic_equation} and
\eqref{angular_geodesic_equation}.  For this purpose, we choose the localized
profile
\begin{equation}\label{gaussian_profile_comparison}
 f(u)=\exp\!\left(-\frac{u^2}{\Lambda^2}\right)
\end{equation}
and use units in which $R=1$, with $A_m=1$, $\Lambda=0.15$,
$u_i=-1$, and $u_f=1$.  The pulse is negligible at both endpoints.  For a
Gaussian profile, the response integral in Eq.~\eqref{response_amplitude} is

\begin{figure}[!t]
 \centering
 \includegraphics[width=\columnwidth]{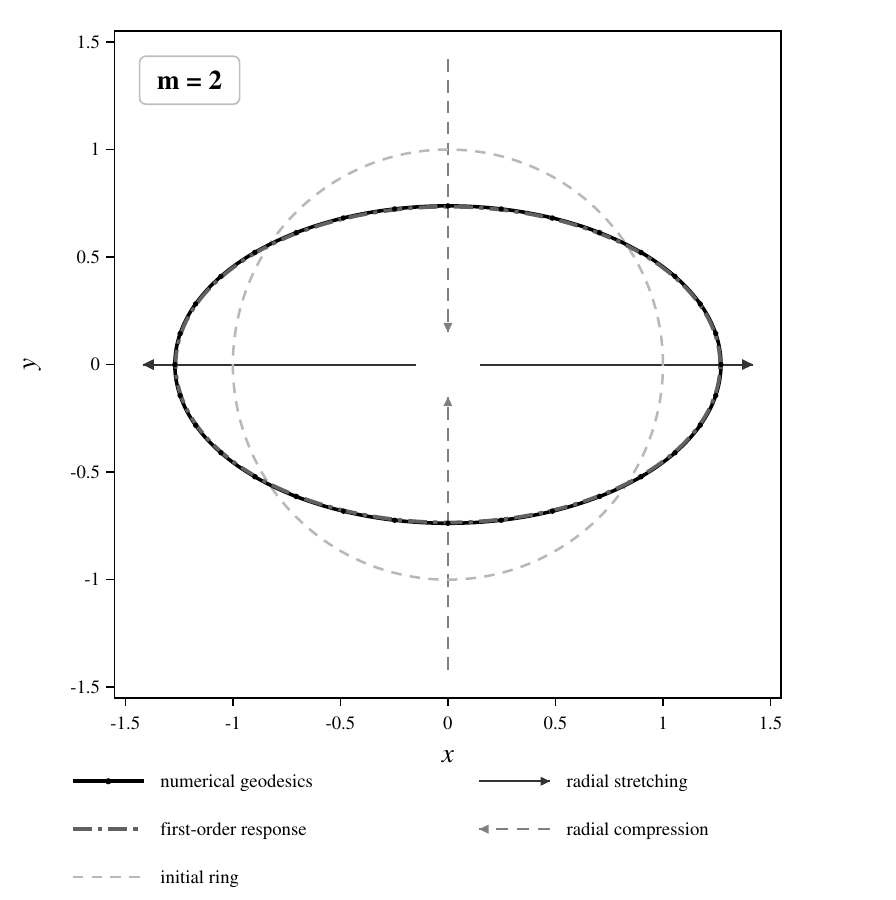}
 \caption{Analytical and numerical deformation for the real $m=2$ mode.
 The solid black curve with markers is obtained by numerical integration of
 the geodesic equations, the gray dash-dotted curve is the first-order
 result~\eqref{first_order_cartesian_ring}, and the light-gray dashed curve
 is the initial ring.  Solid outward arrows mark radial stretching
 directions, whereas dashed inward arrows mark radial compression
 directions.  Here $A_2=1$, $R=1$, $\Lambda=0.15$, $u_i=-1$, and $u_f=1$.}
 \label{fig:floral_m2}
\end{figure}

\begin{figure}[!t]
 \centering
 \includegraphics[width=\columnwidth]{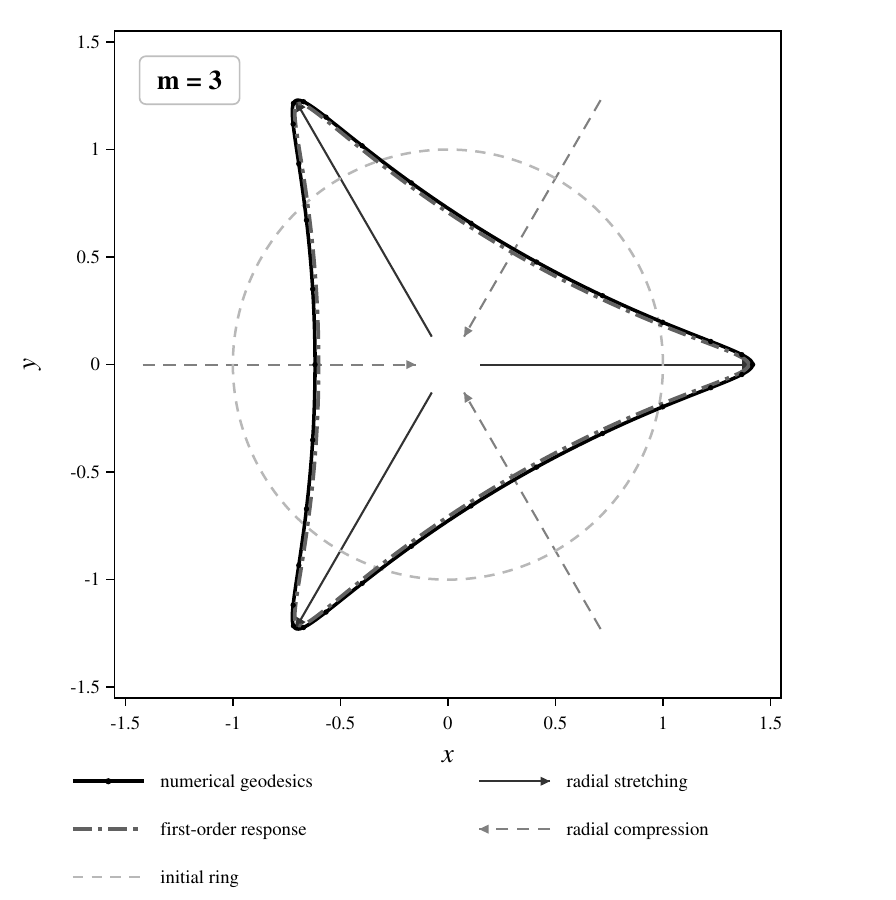}
 \caption{Same comparison as in Fig.~\ref{fig:floral_m2}, now for the real
 $m=3$ mode.  The three radial stretching directions coincide with the
 outward tips of the pattern, while the compression directions bisect
 the angles between neighboring tips.  The analytical curve captures the petal number and
 orientation, with the small difference from the numerical curve measured
 by $\epsilon_3$.}
 \label{fig:floral_m3}
\end{figure}

\begin{figure}[!t]
 \centering
 \includegraphics[width=\columnwidth]{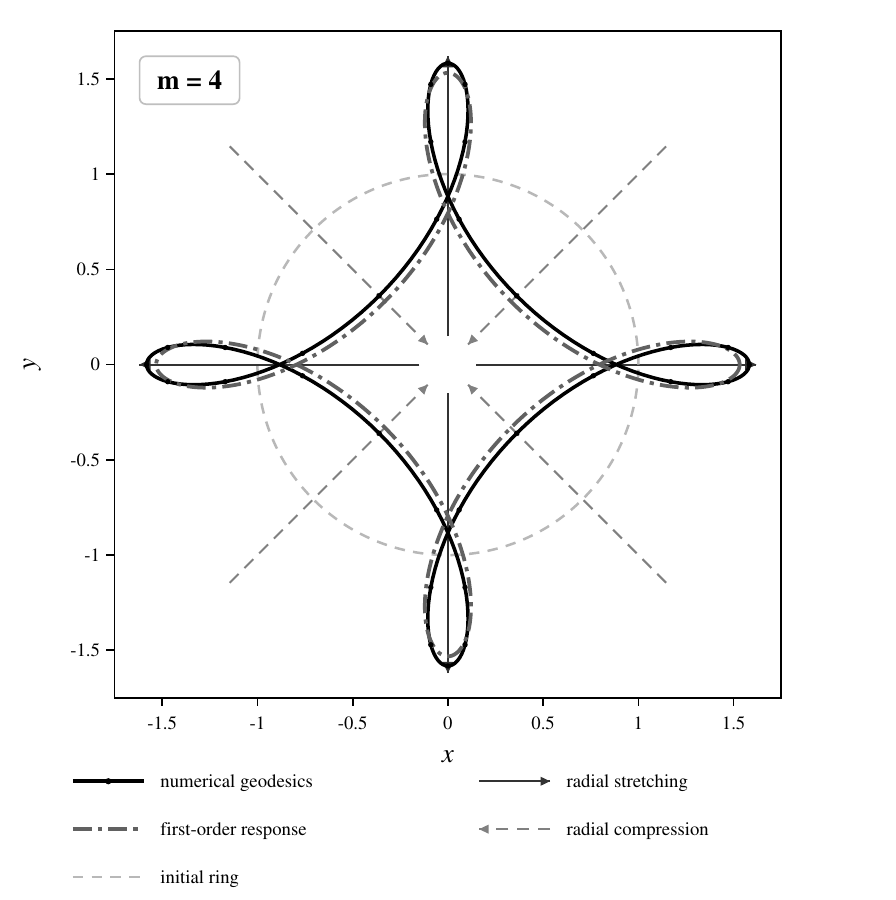}
 \caption{Same comparison as in Fig.~\ref{fig:floral_m2}, now for the real
 $m=4$ mode.  The four stretching directions determine the axes of the four
 loops, and the compression directions lie between them.  The first-order
 curve already predicts the onset of the looped geometry, while its
 separation from the numerical curve is measured by $\epsilon_4$.}
 \label{fig:floral_m4}
\end{figure}

\begin{align}
 {\cal I}(u_f)
 ={}&\frac{\sqrt{\pi}\Lambda u_f}{2}
 \left[
  \operatorname{erf}\!\left(\frac{u_f}{\Lambda}\right)
  -\operatorname{erf}\!\left(\frac{u_i}{\Lambda}\right)
 \right]\nonumber\\
 &+\frac{\Lambda^2}{2}
 \left[
  e^{-u_f^2/\Lambda^2}-e^{-u_i^2/\Lambda^2}
 \right].
 \label{gaussian_response_integral}
\end{align}
For the symmetric interval adopted here,
\begin{equation}
 {\cal I}(1)=\sqrt{\pi}\Lambda
 \operatorname{erf}\!\left(\frac{1}{\Lambda}\right)
 =0.265868\ldots,
\end{equation}
which gives
\begin{equation}\label{comparison_amplitudes}
 {\cal B}_2=0.265868\ldots,
 \qquad
 {\cal B}_3=0.398802\ldots.
\end{equation}
For the quartic mode, the same calculation gives
\begin{equation}\label{comparison_amplitude_m4}
 {\cal B}_4=0.531736\ldots.
\end{equation}

For plotting, Eqs.~\eqref{perturbed_ring_radius} and
\eqref{perturbed_ring_angle} are conveniently combined into the
first-order Cartesian curve
\begin{equation}\label{first_order_cartesian_ring}
 \bm X_m^{(1)}(\theta,u)=
 \begin{pmatrix}
  R\cos\theta+{\cal B}_m(u)
  \cos\!\left[(m-1)\theta-m\varphi_0\right]\\
  R\sin\theta-{\cal B}_m(u)
  \sin\!\left[(m-1)\theta-m\varphi_0\right]
 \end{pmatrix}.
\end{equation}
The numerical ring is constructed from $N=256$ initially equidistant
particles, with no use of the weak-pulse approximation.  We quantify its
deviation from Eq.~\eqref{first_order_cartesian_ring} by
\begin{equation}\label{ring_rms_error}
 \epsilon_m=
 \frac{1}{R}
 \left[
  \frac{1}{N}\sum_{k=0}^{N-1}
  \left\|
   \bm X_{m,k}^{(\mathrm{num})}
   -\bm X_m^{(1)}(\theta_k,u_f)
  \right\|^2
 \right]^{1/2}.
\end{equation}
For the parameters above, the numerical values are
$\epsilon_2=3.83\times10^{-3}$ and
$\epsilon_3=1.73\times10^{-2}$, while
$\epsilon_4=4.63\times10^{-2}$.  The discrepancy therefore grows with the
response amplitude, but the first-order curves still reproduce the
orientation and discrete symmetry of all three numerical patterns.

The $m=4$ result also displays a qualitative transition from simple lobes to
loops.  For the real mode, Eq.~\eqref{first_order_cartesian_ring} can be
written in complex form as
\begin{equation}\label{complex_first_order_ring}
 Z_m^{(1)}(\theta)
 =R e^{i\theta}
 +{\cal B}_m e^{-i[(m-1)\theta-m\varphi_0]}.
\end{equation}
A cusp occurs when $dZ_m^{(1)}/d\theta=0$, which requires
\begin{equation}\label{loop_threshold}
 \frac{|{\cal B}_m|}{R}=\frac{1}{m-1}.
\end{equation}
Above this threshold the curve folds into $m$ symmetry-related loops.  In the
present examples, $\mathcal{B}_3/R=0.399<1/2$, whereas
$\mathcal{B}_4/R=0.532>1/3$.  This explains why the triangular pattern remains
unlooped in Fig.~\ref{fig:floral_m3}, while four loops are already present in
Fig.~\ref{fig:floral_m4}.

The same first-order solution also separates displacement from velocity
memory.  Let the pulse be negligible for $u\geq u_f$ and define its first two
temporal moments by
\begin{equation}\label{pulse_moments}
 F_0=\int_{u_i}^{u_f}f(s)\,ds,
 \qquad
 F_1=\int_{u_i}^{u_f}s f(s)\,ds.
\end{equation}
Writing $K_m=A_m mR^{m-1}/2$, Eq.~\eqref{response_amplitude} then gives, in
the post-pulse region,
\begin{equation}\label{post_pulse_response}
 {\cal B}_m(u)=K_m(uF_0-F_1),
 \qquad
 \dot{\cal B}_m(u)=K_mF_0.
\end{equation}
Thus $F_0\neq0$ produces velocity memory and a floral deformation that keeps
growing linearly, whereas $F_0=0$ and $F_1\neq0$ leave a time-independent
displacement memory.  In either case, Eqs.~\eqref{perturbed_ring_radius} and
\eqref{perturbed_ring_angle} show that the residual response preserves the
same angular order $m$ and orientation $\varphi_0$.  The symmetric Gaussian
used above has $F_1=0$ and $F_0\simeq\sqrt{\pi}\Lambda$, and therefore belongs
to the velocity-memory case.

\subsection{Fourier characterization of the deformed ring}\label{sec3_fourier}

The particle labels provide an unambiguous parametrization even when the
final curve develops loops.  We first remove its centroid,
\begin{equation}\label{centered_final_ring}
 \widetilde{\bm X}_f(\theta_k)=\bm X_f(\theta_k)-\overline{\bm X}_f,
 \qquad
 \overline{\bm X}_f=\frac{1}{N}\sum_{k=0}^{N-1}\bm X_f(\theta_k),
\end{equation}
and project the displacement on the initial polar basis,
\begin{align}
 u_r(\theta)&=
 \left[\widetilde{\bm X}_f(\theta)-\bm X_0(\theta)\right]
 \cdot\widehat{\bm e}_r(\theta),
 \label{radial_displacement_fourier}\\
 u_\varphi(\theta)&=
 \left[\widetilde{\bm X}_f(\theta)-\bm X_0(\theta)\right]
 \cdot\widehat{\bm e}_\varphi(\theta).
 \label{tangential_displacement_fourier}
\end{align}
The radial component is expanded according to
\begin{equation}\label{radial_fourier_series}
 u_r(\theta)=\frac{a_0}{2}
 +\sum_{n=1}^{N/2-1}
 \left[a_n\cos(n\theta)+b_n\sin(n\theta)\right],
\end{equation}
where, for $n\geq1$,
\begin{equation}\label{discrete_fourier_coefficients}
 \begin{split}
  a_n&=\frac{2}{N}\sum_{k=0}^{N-1}u_r(\theta_k)\cos(n\theta_k),\\
  b_n&=\frac{2}{N}\sum_{k=0}^{N-1}u_r(\theta_k)\sin(n\theta_k).
 \end{split}
\end{equation}
We use the amplitude, phase estimator, and spectral purity
\begin{align}
 C_n&=\sqrt{a_n^2+b_n^2},
 \label{fourier_amplitude}\\
 \widehat{\varphi}_0&=\frac{1}{m}\operatorname{atan2}(b_m,a_m),
 \label{fourier_phase}\\
 \Pi_m&=\frac{C_m^2}{\sum_{n=1}^{N/2-1}C_n^2}.
 \label{fourier_purity}
\end{align}

At first order, Eqs.~\eqref{perturbed_ring_radius} and
\eqref{perturbed_ring_angle} give
\begin{equation}\label{first_order_fourier_response}
 u_r^{(1)}={\cal B}_m\cos[m(\theta-\varphi_0)],
 \qquad
 u_\varphi^{(1)}=-{\cal B}_m\sin[m(\theta-\varphi_0)].
\end{equation}
Hence $C_m=|{\cal B}_m|$, $\widehat{\varphi}_0=\varphi_0$ for
${\cal B}_m>0$, and $\Pi_m=1$.  More generally, the exact $m$-fold symmetry
implies the selection rule $C_n=0$ unless $n=qm$, with $q$ a positive integer;
nonlinear evolution can therefore generate $2m,3m,\ldots$, as well as the
uniform term $a_0/2$, but cannot populate arbitrary angular orders.

For the numerical rings of Figs.~\ref{fig:floral_m2}--\ref{fig:floral_m4}, we
obtain
\begin{equation}\label{numerical_fourier_results}
 \begin{array}{c|ccc}
  m & 2 & 3 & 4\\ \hline
  C_m/R
  &0.265888&0.399442&0.535830\\
  C_{2m}/C_m
  &<10^{-12}&6.65\!\times\!10^{-6}&2.02\!\times\!10^{-4}\\
  1-\Pi_m
  &<10^{-12}&4.42\!\times\!10^{-11}&4.08\!\times\!10^{-8}
 \end{array}
\end{equation}
The recovered phases agree with $\varphi_0=0$ to numerical precision.  The
mean radial shifts are $(a_0/2)/R=3.83\times10^{-3}$,
$1.73\times10^{-2}$, and $4.62\times10^{-2}$ for $m=2,3,4$, respectively,
and account for most of the deviations measured by $\epsilon_m$.  Thus, even
the looped $m=4$ curve remains spectrally almost pure: its loops are primarily
the geometrical folding of the fundamental mode above
Eq.~\eqref{loop_threshold}, rather than the result of strong harmonic mixing.

\begin{figure}[!t]
 \centering
 \includegraphics[width=\columnwidth]{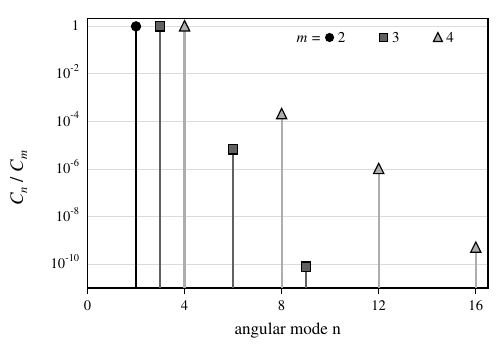}
 \caption{Normalized radial Fourier spectra of the numerical rings, computed
 from centroid-subtracted displacements projected on the initial radial
 directions.  Only $C_n/C_m\geq10^{-11}$ are displayed.  The dominant peak is
 $n=m$, and the nonlinear corrections obey $n=qm$.  Parameters are as in
 Figs.~\ref{fig:floral_m2}--\ref{fig:floral_m4}.}
 \label{fig:fourier_spectra}
\end{figure}

\section{Conclusions}\label{conc}

We have provided an analytical explanation for the floral deformation of
particle rings generated by higher-order polynomial pp-wave modes.  The tidal
eigendirections identify $m$ interlaced sectors of radial stretching and
compression, and the weak-pulse solution shows directly that the ring acquires
$m$-fold symmetry with orientation fixed by $\varphi_0$.  The post-pulse
amplitude is controlled by the first two temporal moments of the profile,
which separate displacement from velocity memory while preserving the same
angular structure.

Direct geodesic integration for $m=2,3,4$ confirms these predictions.  The
Fourier spectra are dominated by $n=m$, with only very small harmonics at
$n=qm$; the mean radial shift accounts for most of the departure from the
first-order curves.  In particular, the four loops of the $m=4$ pattern result
primarily from the geometrical folding of an almost pure fundamental mode once
$|{\cal B}_m|/R>1/(m-1)$, rather than from strong nonlinear mode mixing.  The
local tidal geometry, global ring morphology, and residual memory are thus
different manifestations of the same mode-dependent angular response.


\section*{Acknowledgements}

The development of this research was supported by the Research and Graduate Support Program (PROPESQ/UFNT), under Grant No.~030/2026. F. L. Carneiro acknowledges financial support from the Brazilian National Council for Scientific and Technological Development (CNPq), Brazil, through a Productivity Fellowship (PQ), Grant No. 303574/2026-7.


\bibliographystyle{unsrt}
\bibliography{bibitex}

@article{christodoulou1991nonlinear,
  title   = {Nonlinear Nature of Gravitation and Gravitational-Wave Experiments},
  author  = {Christodoulou, D.},
  journal = {Physical Review Letters},
  volume  = {67},
  number  = {12},
  pages   = {1486--1489},
  year    = {1991},
  doi     = {10.1103/PhysRevLett.67.1486}
}

@article{thorne1992gravitational,
  title   = {Gravitational-Wave Bursts with Memory: The Christodoulou Effect},
  author  = {Thorne, K. S.},
  journal = {Physical Review D},
  volume  = {45},
  number  = {2},
  pages   = {520--524},
  year    = {1992},
  doi     = {10.1103/PhysRevD.45.520}
}

@article{podolsky2014gyratonic,
  title={Gyratonic pp waves and their impulsive limit},
  author={Podolsk{\`y}, J. and Steinbauer, R. and {\v{S}}varc, R.},
  journal={Physical Review D},
  volume={90},
  number={4},
  pages={044050},
  year={2014},
  publisher={APS}
}

@article{zhang2018memory,
  title={Memory effect for impulsive gravitational waves},
  author={Zhang, P. M. and Duval, C. and Horvathy, P. A.},
  journal={Classical and Quantum Gravity},
  volume={35},
  number={6},
  pages={065011},
  year={2018},
  publisher={IOP Publishing}
}

@article{zhang2024displacement,
  title={Displacement within velocity effect in gravitational wave memory},
  author={Zhang, P. M. and Horvathy, P. A.},
  journal={Annals of Physics},
  volume={470},
  pages={169784},
  year={2024},
  publisher={Elsevier}
}

@article{zhang2017memory,
  title={The memory effect for plane gravitational waves},
  author={Zhang, P. M. and Duval, C. and Gibbons, G. W. and Horvathy, P. A.},
  journal={Physics Letters B},
  volume={772},
  pages={743--746},
  year={2017},
  publisher={Elsevier}
}

@article{maluf2018plane,
  title={Plane gravitational waves, the kinetic energy of free particles and the memory effect},
  author={Maluf, J. W. and Rocha-Neto, J. F. and Ulhoa, S. C. and Carneiro, F. L.},
  journal={Gravitation and Cosmology},
  volume={24},
  number={3},
  pages={261--266},
  year={2018},
  publisher={Springer}
}

@article{zhao2026plane,
  title={Plane-wave memory beyond the sandwich approximation},
  author={Zhao, Q. L. and Cao, L. M.},
  journal={Physics Letters B},
  pages={140783},
  year={2026},
  publisher={Elsevier}
}

@article{datta2024memory,
  title={Memory effect of gravitational wave pulses in pp--wave spacetimes},
  author={Datta, S. and Guha, S.},
  journal={Physica Scripta},
  volume={99},
  number={7},
  pages={075023},
  year={2024},
  publisher={IOP Publishing}
}

@article{zhang2025displacement,
  title={Displacement memory for flyby},
  author={Zhang, P. M. and Zhao, Q. L. and Balog, J. and Horvathy, P. A.},
  journal={Annals of Physics},
  volume={473},
  pages={169890},
  year={2025},
  publisher={Elsevier}
}

@article{elbistan2026globally,
  title={Globally defined Carroll symmetry of gravitational waves},
  author={Elbistan, M. and Zhang, P. M. and Horvathy, P. A.},
  journal={Nuclear Physics B},
  pages={117354},
  year={2026},
  publisher={Elsevier}
}

@article{zhang2025flyby,
  title={Flyby-induced displacement effect: An analytic solution},
  author={Zhang, P. M. and Silagadze, Z. K. and Horvathy, P .A.},
  journal={Physics Letters B},
  volume={868},
  pages={139687},
  year={2025},
  publisher={Elsevier}
}

@article{achour2024displacement,
  title={Displacement versus velocity memory effects from a gravitational plane wave},
  author={Achour, J. B. and Uzan, J. P.},
  journal={Journal of Cosmology and Astroparticle Physics},
  volume={2024},
  number={08},
  pages={004},
  year={2024},
  publisher={IOP Publishing}
}

@article{zhao2026decay,
  title={Decay criteria for asymptotic freedom in plane gravitational waves},
  author={Zhao, Q. L. and Cao, L. M.},
  journal={arXiv preprint arXiv:2605.29636},
  year={2026}
}

@article{zhang2026memory,
  title={Memory Effect for deformed gravitational waves},
  author={Zhang, P. M. and Elbistan, M. and Horvathy, P. A.},
  journal={arXiv preprint arXiv:2607.25452},
  year={2026}
}

@article{carneiro2026memory,
  title={Memory effect for generalized modes in pp-waves spacetime},
  author={Carneiro, F. L. and de Carvalho, H. P. and Lobo, M. P. and Cabral, L. A.},
  journal={arXiv preprint arXiv:2603.27042},
  year={2026}
}

@article{rossetto2026wada,
  title={Wada Boundaries in Generic Polynomial pp--Wave Spacetimes},
  author={Rossetto, P. H. B. and de Andrade, V. C. and M{\"u}ller, D.},
  journal={arXiv preprint arXiv:2601.09101},
  year={2026}
}

@article{zhao2026approximate,
  title={Approximate models for gravitational memory},
  author={Zhao, Q. L. and Zhang, P. M. and Elbistan, M. and Horvathy, P. A.},
  journal={Physics Letters B},
  volume={879},
  pages={140603},
  year={2026},
  publisher={Elsevier}
}

\end{document}